\documentclass[usenatbib]{arxiv}

\usepackage{newtxtext,newtxmath}

\usepackage[T1]{fontenc}

\DeclareRobustCommand{\VAN}[3]{#2}
\let\VANthebibliography\thebibliography
\def\thebibliography{\DeclareRobustCommand{\VAN}[3]{##3}\VANthebibliography}

\usepackage{graphicx}	
\usepackage{amsmath}	
\allowdisplaybreaks     

\usepackage{bm}
\usepackage{comment}

\newcommand{\bvec}[1]{\bm{#1}}
\newcommand{\unid}[1]{\ensuremath{\:\mathrm{#1}}}

\graphicspath{{Imagens/}}

\title[General-relativistic radiative cooling]{General-relativistic and non-ideal radiative cooling in neutron star magnetospheres}

\author[J. Joaquim et al.]{
J. Joaquim,$^{1}$\thanks{E-mail: joao.s.joaquim@tecnico.ulisboa.pt}
F. Assunção,$^{1}$
P. J. Bilbao,$^{1,2}$
L. O. Silva$^{1}$\thanks{E-mail: luis.silva@tecnico.ulisboa.pt}
\\
$^{1}$GoLP/Instituto de Plasmas e Fusão Nuclear, Instituto Superior Técnico, Universidade de Lisboa, 1049-001 Lisboa, Portugal\\
$^{2}$The Rudolf Peierls Centre for Theoretical Physics, University of Oxford, Oxford OX1 3NP, United Kingdom\\
}

\pubyear{\the\year{}}

\begin{document}
\label{firstpage}
\pagerange{\pageref{firstpage}--\pageref{lastpage}}
\maketitle

\begin{abstract}

Radiation reaction cooling plays an important role in describing the extreme plasma conditions found in the magnetospheres of astrophysical compact objects. Strong electromagnetic fields, characteristic of these environments, can trigger the development of anisotropic ring-shaped plasma distributions with inverted Landau populations in momentum space. In this work, we present the first systematic investigation of this mechanism in realistic astrophysical configurations, by accounting for how non-uniform electromagnetic field geometries and general-relativistic effects modify the phase-space dynamics of radiatively cooled plasmas. We demonstrate analytically that drift velocities favour the formation of spiral-shaped momentum distributions that still display inverted Landau populations, and estimate the minimum and maximum plasma injection distances required for inverted momentum distributions to be able to power the emission of coherent radiation through kinetic instabilities. From numerical simulations, we conclude that curved spacetime increases the gradient of the distribution function responsible for the development of kinetic instabilities, and prolongs the persistence of the inverted momentum structure relative to flat spacetime, confirming that realistic astrophysical conditions preserve and enhance the conditions necessary for synchrotron-powered emission of coherent radiation to occur.

\end{abstract}



\section{Introduction}

Astrophysical compact objects such as neutron stars and black holes power some of the most extreme physical phenomena in the universe.
Among these, rotating magnetized neutron stars, such as pulsars and magnetars, possess magnetospheres that harbor ultra-strong gravitational and electromagnetic fields \citep{nsbfieldestimate,goldreichjulian,nsbfields1,magnetarbfield,teukolskycompactobjects}, thereby creating an environment in which general-relativistic and radiative processes couple to the collective
plasma dynamics \citep{philippovabinitio,RRastro2,limitsextremeplasma}.

Such extreme magnetospheric conditions are at the core of the generation of coherent radiation in neutron stars, including pulsar radio emission and fast radio bursts (FRBs) \citep{pulsardiscovery,gunn1970nature,coherenttemperature2,frbdiscovery,frbreview,magnetospherereview,FRBmagnetars2023}.
However, despite decades of theoretical work and multiple proposed models \citep{coherentemissionmechanisms2,coherentemissionmechanisms3,shockdrivenmasers1,shockdrivenmasers2,frborigincompactobj1,freeelectrons1,FRBalign} (for a review, see \cite{coherentemissionmechanisms1}), the origin of astrophysical coherent emission in neutron stars remains elusive and one of the most outstanding problems in high-energy astrophysics.

Recent works \citep{RRcoolingflat3,pablorings2023,pablorings2024,ochs2024synchrotron} showed that radiation reaction (RR) cooling compresses the phase-space volume and reduces the entropy of magnetized plasmas, thereby enabling the development of non-thermal ring-shaped momentum distribution functions (MDFs). These distributions are anisotropic and develop inverted Landau populations in finite time, and can thus act as a source of free energy for kinetic instabilities capable of producing coherent emission, such as the electron cyclotron maser instability (ECMI), whose emission properties are consistent with several defining characteristics of neutron star coherent emission \citep{pabloECMIrings}. 
These studies have been carried out in flat spacetime, under idealized field configurations.  

In this work, we study how non-uniform electromagnetic field geometries and curved spacetime effects modify the phase-space dynamics of radiatively cooled plasmas.
We extend the formalism of radiation reaction cooling to curved spacetime and perform numerical simulations of ensembles of charged test particles in neutron star magnetospheres, resorting to a parallelized particle pusher that integrates particle trajectories in stationary electromagnetic fields and background spacetime metrics.
We aim to provide a first picture of how the formation of inverted MDFs manifests itself in realistic astrophysical contexts. 

This paper is organized as follows. In Section \ref{sec:2}, we establish the theoretical formalism underpinning the work. We extend the radiative Vlasov equation to the 3+1 formalism of general relativity (GR) and examine the astrophysical conditions necessary for coherent emission due to synchrotron-induced ECMI to occur. In Section \ref{sec:3}, we demonstrate how the effects of non-uniform electromagnetic field geometries favor the development of spiral-shaped momentum distributions, which display inverted Landau populations and are thus kinetically unstable. In Section \ref{sec:4}, we perform a systematic study of radiative cooling in realistic astrophysical configurations. Resorting to both rescaled and realistic parameters, we analyze the formation, evolution, and properties of spiral distributions in neutron star spacetimes, and how they affect ECMI emission as compared with flat spacetime. Finally, in Section \ref{sec:conclusions}, we discuss our key findings.

In this paper, we employ geometrized units $G = c = 1$ when working with general-relativistic quantities, and the metric signature $(-,+,+,+)$.
Greek indices run from 0 to 3 and denote four-dimensional quantities, while Latin indices run from 1 to 3 and designate three-dimensional quantities. The Einstein summation convention is employed unless explicitly stated otherwise.
All physical quantities are presented in the CGS system of units.

\section{Radiative cooling in curved spacetime}
\label{sec:2}

\subsection{Neutron star spacetimes}
\label{subsec:kerrslowmetric}

General relativity is a key ingredient in describing the dynamics of plasmas in neutron star magnetospheres \citep{pulsarmagGR,petrimagGR}. The external spacetime of rotating neutron stars is described by the Hartle-Thorne metric \citep{hartlethorne2}, which represents the exterior spacetime of a rotating, stationary, axisymmetric massive body to second order in the angular frequency $\Omega_*$.
It is often sufficient to employ its slow-rotation limit $R_* \Omega_* \ll 1$, where $R_*$ is the stellar radius, which coincides exactly with the first-order expanded Kerr metric \citep{kerr1963}.
The resulting approximated metric is known as slowly-rotating Kerr, or Kerr-slow (KS), and its line element is determined in Boyer-Lindquist coordinates $(t,r,\theta,\phi)$ by

\begin{align}
    \label{eq:ds2KS}
     ds^2 &= -\left( 1-\frac{r_s}{r} \right) dt^2 + \left( 1-\frac{r_s}{r} \right)^{-1} dr^2 \\ &+ r^2 (d\theta^2 + \sin^2{\theta} \ d\phi^2) - 2\omega_{LT}(r)r^2\sin^2{\theta} \ dt d\phi \ \nonumber .
\end{align}

\noindent
where $r_s = 2M$ is the Schwarzschild radius of the star, and is always smaller than $R_*$ in neutron stars. 

In rotating geometries, spacetime is dragged along the rotating mass, and bodies initially at rest acquire rotational motion of frequency $\omega_{LT}(r)\approx 0.21\,\Omega_*\frac{r_s}{R_*-r_s} \left(\frac{R_*}{r}\right)^3$ \citep{NSmomentsofinertia}, an effect known as the Lense-Thirring effect, or frame-dragging (FD).

In this work, we employ the slowly rotating Kerr metric, which provides a spacetime background that captures the effects of rotation while remaining analytically and computationally manageable. The line element (\ref{eq:ds2KS}) allows the determination of the necessary quantities for applying the 3+1 formalism of GR \citep{admformalism,gourgoulhon3+1} to Kerr-slow spacetime, which consist of the lapse function $\alpha = (-g^{00})^{-1/2}= \sqrt{1-\frac{r_s}{r}}$, and the shift vector $\beta^i = \alpha^2 g^{0i} = \Big( 0,0,-\omega_{LT}(r) \Big)$, which are uniquely determined from the metric tensor $g_{\mu\nu}$.

\subsection{Radiative Vlasov equation in curved spacetime}
\label{subsec:vlasov3+1} 

The dynamics of the plasmas that permeate the magnetospheres of compact objects are affected by strong general-relativistic effects. Such plasmas can be considered collisionless \citep{aronscollisionless,collisionless1}, and are accurately modelled by the Vlasov-Maxwell system in a curved spacetime background \citep{cowling1,EVMsystem,osirisgr}. 
Still, this description misses the critical role of radiation; to account for radiative dissipation, the Vlasov equation must be extended to include radiation reaction. 

Kinetic equations that include dissipative forces have been extensively studied in flat spacetime \citep{vlasovRR1,vlasovRRentropy1,vlasovRRentropy2}.
Generalizing the procedure to introduce non-conservative force terms in kinetic equations, to the curved spacetime Vlasov equation \citep{GRkintheoryarticle}, we obtain the radiative curved spacetime Vlasov equation:

\begin{equation}
    \label{eq:radiativeEVMvlasov2}
    \frac{p^\mu}{m} \frac{\partial f}{\partial x^\mu} + \left( -\Gamma^\mu_{\alpha \beta} \frac{p^\alpha p^\beta}{m} + \frac{q}{m} F^{\mu\nu} p_\nu + F_{RR}^\mu \right) \frac{\partial f}{\partial p^\mu} = -f \frac{\partial F_{RR}^\mu}{\partial p^\mu} \ ,
\end{equation}

\noindent
which describes a distribution of electrons particles of mass $m$, charge $q$, and 4-momentum $p^\mu = m u^\mu$, subject to an electromagnetic field $F^{\mu\nu}$ in a curved spacetime background with Christoffel symbols $\Gamma^\mu_{\alpha\beta}$. Since this work concerns electron-positron pair plasmas, we define $q$ and $m$ respectively as the electron charge and mass.

Since the Vlasov equation is a first-order partial differential equation, its characteristics can be obtained directly from equation (\ref{eq:radiativeEVMvlasov2}). These are exactly the covariant equations of motion for a charged particle in curved spacetime, and therefore, their 3+1 decomposition yields \citep{3+1EOM2,gourgoulhon3+1}:

\begin{equation}
    \label{eq:3+1eqmom}
     \frac{dx^i}{dt} \bvec{e}_i = \frac{\alpha\bvec{p}}{\gamma m} - \bvec{\beta} \ \ ; \ \ \frac{d \bvec{p}}{dt} = \alpha \left[ \gamma m \bvec{g} + H \cdot \bvec{p} + q \left(\bvec{E} + \frac{\bvec{p}}{\gamma m} \times \bvec{B} \right) + \bvec{F}_{RR} \right] \ ,
\end{equation}

\noindent
where $t$ is the global coordinate time provided by fiducial observers and $\bvec{p}$ is the spatial hypersurface projection of the 4-momentum $p^\mu$. The usage of the 3+1 formalism gives rise to two new quantities: the gravitational acceleration $\bvec{g} = -\bvec{\nabla} \ln{\alpha}$, and the gravitomagnetic tensor $H = \alpha^{-1} \bvec{\nabla} \bvec{\beta}$, which are defined in terms of the lapse function and the shift vector. 

Moreover, $\bvec{F}_{RR}$ represents the classical radiation reaction force, which we model according to the general-relativistic reduced Landau-Lifshitz (LLR) model, whose derivation is detailed in Appendix \ref{app:RRG}. This force is expressed in 3+1 form as:

\begin{equation}
    \label{eq:RRLLcurved}
   \bvec{F}_{RR} = \bvec{F}_{RR}^{(LL)} + \bvec{F}_{RR}^{(G)} \ ,  
\end{equation}

\noindent
where $\bvec{F}_{RR}^{(LL)}$ is the reduced Landau-Lifshitz force \citep{landaulifshitzbook}, and $\bvec{F}_{RR}^{(G)}$ is the curved spacetime contribution, which is given by \citep{jjthesis}:

\begin{align}
    \label{eq:LLRR3D}
     \bvec{F}_{RR}^{(LL)} = \frac{2q^4}{3m^2} &\Bigg[ \left(\bvec{E} + \frac{\bvec{p}}{\gamma m} \times \bvec{B}\right) \times \bvec{B} + \bvec{E}\left(\frac{\bvec{p}}{\gamma m} \cdot \bvec{E}\right) \\ & \ \ -\frac{\gamma \bvec{p}}{m} \left(\left(\bvec{E} + \frac{\bvec{p}}{\gamma m} \times \bvec{B}\right)^2 - \left(\frac{\bvec{p}}{\gamma m} \cdot \bvec{E}\right)^2 \right) \Bigg] \nonumber \ ,
\end{align}

\begin{align}
    \label{eq:LLRRG3D}
    F_{RR}^{(G) i} = \frac{2q^3 \gamma}{3m} &\Bigg[ \Gamma^i_{jk} \frac{p^j}{\gamma m} \left(\bvec{E}+\frac{\bvec{p}}{\gamma m}\times \bvec{B}\right)^k - \Gamma^j_{kl} \frac{p^k p^l}{\gamma^2 m^2} \epsilon^i_{ \ jm} B^m \\ & \ \ + (\bvec{g} \times \bvec{B})^i + \left((\bvec{g} \times \bvec{E}) \times \frac{\bvec{p}}{\gamma m}\right)^i + \left(\bvec{g} \cdot \frac{\bvec{p}}{\gamma m}\right)E^i \Bigg] \ \nonumber .
\end{align}

\noindent
In the latter, $\Gamma^i_{jk}$ are the three-dimensional Christoffel symbols and $\epsilon_{\mu\nu\rho\sigma} = \sqrt{-\det{g_{\alpha\beta}}} \,  \varepsilon_{\mu\nu\rho\sigma}$ is the Levi-Civita tensor.
We note that equation (\ref{eq:LLRRG3D}) is coordinate-dependent, and was derived for the special case of stationary spacetimes expressed in Boyer-Lindquist coordinates, whose only non-zero diagonal entries are $g_{t\phi}$.
Schwarzschild and Kerr spacetimes are examples of such metrics.

Beyond the classical formalism discussed above, quantum corrections to radiation reaction become significant when the quantum non-linearity parameter $\chi\approx\gamma B/B_{sc}$ surpasses unity \citep{quantumRRparam1,quantumRRparam2,marijaquantumRR}. In the worst-case scenario of our simulation parameter regime, $\chi = 1$ only at the initialization instant, and $\chi < 1$ for the remainder of the simulation. As the distribution cools down, $\chi \rightarrow 0$ and quantum radiative effects very rapidly become dominated by the classical force. In previous works \citep{pablorings2023,pablorings2024}, it was shown that the main difference introduced by the quantum regime is the phase-space diffusion resulting from QED synchrotron emission, and that population inversion is still produced within comparable timescales as in the classical regime. Therefore, we expect that the mechanisms discussed here remain relevant, and qualitatively apply, also in the quantum regime $\chi \gtrsim 1$.

\medskip

The Vlasov equation in 3+1 form then reads:

\begin{align}
    \label{eq:3+1radvlasov}
    &\frac{\partial f}{\partial t} + \left(\frac{\alpha\bvec{p}}{\gamma m} - \bvec{\beta} \right) \cdot \bvec{\nabla_x} f +  \alpha \bigg[\gamma m \bvec{g} + H \cdot \bvec{p} \\ & + q \left(\bvec{E} + \frac{\bvec{p}}{\gamma m} \times \bvec{B} \right) + \bvec{F}_{RR} \bigg] \cdot \bvec{\nabla_p} f = -\alpha f \bvec{\nabla_p} \cdot \bvec{F}_{RR} \ \nonumber . 
\end{align}

\noindent
In the absence of radiative losses and gravity, $\alpha=1$ and $\bvec{\beta}=\bvec{F}_{RR}=0$, and equation (\ref{eq:3+1radvlasov}) reduces to the usual flat-spacetime Vlasov equation, as expected.

\section{Synchrotron-induced coherent emission in realistic magnetospheric configurations}
\label{sec:3}

The mechanism through which radiation reaction cooling leads to the formation of inverted momentum distributions that trigger the electron cyclotron maser instability offers a first-principles model for the generation of coherent radio emission in the magnetospheres of neutron stars. In these astrophysical environments, the formation and evolution of ring distributions is governed by the general-relativistic Vlasov equation (\ref{eq:3+1radvlasov}), which includes gravitational acceleration, frame-dragging, and the effects of electric fields and non-uniform field geometries. These contributions can affect radiative cooling and will thus have an impact on the phase-space structure.

\subsection{Spiral momentum distributions due to drift velocities}
\label{subsec:spirals} 

Before considering the full effect of realistic astrophysical configurations, we perform an analytical extension of the works of \cite{pablorings2023} and \cite{pablorings2024} to assess the impact of a general drift velocity, $\bvec{v}_d$, on the radiative cooling process of a plasma.

We consider only scenarios where charges moving in an otherwise strong uniform magnetic field are perturbed by some inhomogeneity.
Provided that the time and length scales governing the variation of this inhomogeneity, $T \equiv | \partial / \partial t |^{-1}$ and $L \equiv | \partial / \partial \bvec{x} |^{-1}$, respectively, are much larger than the cyclotron period, $T \gg T_L = 2 \pi \gamma / \omega_c$, and the Larmor radius of the plasma particles, $L \gg \rho_L = \gamma v_\perp / \omega_c$, where $\omega_c = q B / (m c)$ is the cyclotron frequency, then their motion is approximately described as the sum of the usual gyration motion around the B-field lines with a small drift velocity of their guiding center, $| \bvec{v}_d | / v_\perp  \sim \mathcal{O} (\rho_L / L)$, perpendicular to $\bvec{B}$.
The particle momentum is thus split as $\bvec{p} \simeq \bvec{p}_\parallel + \bvec{p}' \vphantom{p}_{\!\!\perp} + \gamma m \bvec{v}_d$, where $\bvec{p}_\parallel = ( \bvec{p} \cdot \bvec{e}_\parallel ) \bvec{e}_\parallel$ is the momentum component parallel to the magnetic field lines, $\bvec{e}_\parallel \equiv \bvec{B}/B$, $\bvec{p}' \vphantom{p}_{\!\!\perp}$ describes the gyration motion of the particles in the plane perpendicular to $\bvec{B}$, and $\gamma = \sqrt{1 + \bvec{p} \cdot \bvec{p}/(m c)^2}$ is the Lorentz factor.
These conditions are easily fulfilled in the strongly magnetized environments we are considering, as $\omega_{c}$ and $\rho_L$ are usually far removed from the other characteristic scales of the system.

For the sake of generality, we refrain from specifying the origin of this drift velocity in this section.
Possible examples could be the usual particle drifts arising from a guiding center approximation in non-uniform electromagnetic field geometries \citep{Vandervoort,Northrop}, or from the presence of gravitational forces.
Under these assumptions and following \cite{Li}, an approximate solution of the Landau-Lifshitz equations \citep{landaulifshitzbook} for particle motion is (ignoring gravitational time dilation, $\alpha \sim 1$)
\begin{subequations}\label{eq:approximate_motion_solution}
\begin{align}
\begin{split}
p_{\perp 1} &= \frac{m c}{p' \vphantom{p}_{\!\!\perp 0}} \frac{p_{\perp 1 0} \cos \psi - p' \vphantom{p}_{\!\!\perp 2 0} \sin \psi}{\sinh \Big( \frac{\gamma' \vphantom{p}_{\!\!\perp 0}}{\gamma' \vphantom{p}_{\!0}} \omega_{RR} t + \textrm{arsinh} \big( \frac{m c}{p' \vphantom{p}_{\!\!\perp 0}} \big) \Big)} \ ,
\end{split}
\\[2pt]
\begin{split}
p' \vphantom{p}_{\!\!\perp 2} &=  \frac{m c}{p' \vphantom{p}_{\!\!\perp 0}} \frac{p' \vphantom{p}_{\!\!\perp 2 0} \cos \psi + p_{\perp 1 0} \sin \psi}{\sinh \Big( \frac{\gamma' \vphantom{p}_{\!\!\perp 0}}{\gamma' \vphantom{p}_{\!0}} \omega_{RR} t + \textrm{arsinh} \big( \frac{m c}{p' \vphantom{p}_{\!\!\perp 0}} \big) \Big)} \ ,
\end{split}
\\[2pt]
\begin{split}
p_\parallel &= \frac{1}{\gamma' \vphantom{p}_{\!\!\perp 0}} \frac{p_{\parallel 0}}{\tanh \Big( \frac{\gamma' \vphantom{p}_{\!\!\perp 0}}{\gamma' \vphantom{p}_{\!0}} \omega_{RR} t + \textrm{artanh} \big( \frac{1}{\gamma' \vphantom{p}_{\!\!\perp 0}} \big) \Big)} \ ,
\end{split}
\\[2pt]
\begin{split}
\gamma' &= \frac{1}{\gamma' \vphantom{p}_{\!\!\perp 0}} \frac{\gamma'_0}{\tanh \Big( \frac{\gamma' \vphantom{p}_{\!\!\perp 0}}{\gamma' \vphantom{p}_{\!0}} \omega_{RR} t + \textrm{artanh} \big( \frac{1}{\gamma' \vphantom{p}_{\!\!\perp 0}} \big) \Big)} \ ,
\end{split}
\end{align}
\end{subequations}
where $p_{\perp 1}$ and $p_{\perp 2}$ are the two momentum components perpendicular to the magnetic field lines, $\bvec{e}_\parallel \perp \bvec{e}_{\perp 1}$ and $\bvec{e}_{\perp 2} \equiv \bvec{e}_\parallel \times \bvec{e}_{\perp 1}$, the $0$ subscript denotes the initial condition of the quantities, and the primed quantities, $p' \vphantom{p}_{\!\!\perp 2} = \gamma_d (p_{\perp 2} - \gamma m v_d)$ and $\gamma' = \gamma_d (\gamma - p_{\perp 2} v_d /(mc^2))$, with $\gamma_d = 1/\sqrt{1 - (v_d/c)^2}$, are Lorentz-boosted to the reference frame comoving with the drift velocity, which we have assumed to be oriented along $\bvec{e}_{\perp 2}$.
Note that the solution for $p_{\perp 2}$ can be obtained by employing the inverse Lorentz transform, $p_{\perp 2} = \gamma_d (p' \vphantom{p}_{\!\!\perp 2} + \gamma' m v_d)$. 
We also introduce the quantities $p' \vphantom{p}_{\!\!\perp} \vphantom{p}^{\!\!2} \equiv p \vphantom{p}_{\perp 1} \vphantom{p}^{\!\!2} + p' \vphantom{p}_{\!\!\!\perp 2} \vphantom{p}^{\!\!\!2}$ and $\gamma' \vphantom{p}_{\!\!\perp} \vphantom{p}^{\!\!2} \equiv \gamma'^{2} - p_\parallel \vphantom{p}^{\!2}$, and the gyrophase of the particle is given by
\begin{equation}\label{eq:gyrophase}
\!\!\!\psi = \frac{\omega_c}{\omega_{RR}} \! \ln \!\!\, \left[ \frac{1}{\gamma' \vphantom{p}_{\!\!\perp 0}} \frac{p' \vphantom{p}_{\!\!\perp 0}}{m c} \cosh \! \left( \frac{\gamma' \vphantom{p}_{\!\!\perp 0}}{\gamma' \vphantom{p}_{\!0}} \omega_{RR} t + \textrm{arcosh} \! \left( \gamma' \vphantom{p}_{\!\!\perp 0} \frac{m c}{p' \vphantom{p}_{\!\!\perp 0}} \right) \right) \right] \! ,
\end{equation}
where $\omega_{RR} = (2/3) \alpha_S (B/B_{sc}) \omega_c$ is a characteristic frequency for radiation damping, with $\alpha_S = q^2/\hbar c$ being the fine-structure constant and $B_{sc} = m^2 c^3/(q \hbar) \simeq 4.41 \times 10^{13} \ \textrm{G}$ the Schwinger magnetic field \citep{schwingerlimit}.

In the absence of a drift velocity, $\bvec{v}_d \rightarrow 0$, we recover equations (5) and (6) from \cite{pablorings2024}.
When we consider a small drift velocity, $\gamma_d \sim 1$, the guiding center of the particle shifts in momentum space to the point $\gamma m \bvec{v}_d$ and, consequently, the rate for energy loss through radiation emission becomes a function of the perpendicular momentum of the particle, only now with respect to its new guiding center, i.e., it will depend on the Lorentz-boosted perpendicular momentum, $\bvec{p}' \vphantom{p}_{\!\!\perp} \simeq \bvec{p}_\perp - \gamma m \bvec{v}_d$.
Moreover, for particles along the phase-space curve given by $p_{\perp 1} = 0$ and $p_{\perp 2} = \gamma_d m v_d \sqrt{1 + p_\parallel \vphantom{p}^{\!2} /(mc)^2}$, we have $p' \vphantom{p}_{\!\!\perp} = 0$ and $\gamma' \vphantom{p}_{\!\!\perp} = 1$, and it follows from equations \eqref{eq:approximate_motion_solution} that their velocity will remain constant and equal to $\bvec{v} = \bvec{v}_\parallel + \bvec{v}_d$, i.e., they do not radiate.
This implies that all particles in momentum space will gyrate around this curve and, as $t \rightarrow \infty$, their trajectories converge to a point somewhere along it.
\\

Finding how the plasma MDF evolves entails computing a solution of the Vlasov equation \eqref{eq:3+1radvlasov} which, for the simplified scenario considered here, reduces to
\begin{align}\label{eq:vlasov_drift}
\begin{split}
\frac{\partial f}{\partial t} + \frac{\omega_c}{\gamma} &\Bigg[ p' \vphantom{p}_{\!\!\perp 2} \frac{\partial f}{\partial p_{\perp 1}} - p_{\perp 1} \frac{\partial f}{\partial p' \vphantom{p}_{\!\!\perp 2}} \Bigg]
\\
= \frac{\omega_{RR}}{\gamma} &\Bigg[ p_\parallel \frac{p'_\perp \vphantom{p}^{\!\!2}}{(m c)^2} \frac{\partial f}{\partial p_\parallel} + p'_\perp \gamma'_\perp \vphantom{p}^{\!\!2} \frac{\partial f}{\partial p'_\perp} + \left( 4 \gamma'_\perp \vphantom{p}^{\!\!2} - 2 \right) f \Bigg] \ ,
\end{split}
\end{align}
where we have neglected the spatial gradient of the distribution function, $\boldsymbol{\nabla}_{\bvec{x}} f \rightarrow 0$, since we are studying local kinetic properties of the plasma and thus consider it to be homogeneous.
From the solution for single-particle motion, we can then retrieve the temporal evolution of the plasma MDF given its initial state, $f_0(\bvec{p}_0) \equiv f(\bvec{p},t=0)$, and build some intuition on what types of structures can form in momentum space in the presence of a drift velocity.
If we consider a small volume in momentum space, $V(t)$, the number of particles it contains must be conserved at all times as it deforms due to the dissipative nature of radiation reaction, that is,
\begin{equation}
N = \iiint_{V(t)} f(\bvec{p},t) d \bvec{p} = \iiint_{V_0} f_0(\bvec{p}_0) d \bvec{p}_0 ,
\end{equation}
implying that $f(\bvec{p},t) = f_0 ( \bvec{p}_0 ) | \partial \bvec{p}_0/\partial \bvec{p}|$.
By inverting equations \eqref{eq:approximate_motion_solution}, we obtain $\bvec{p}_0 (\bvec{p},t)$, written in full in Appendix \ref{app:math_section3}, that maps $\{\bvec{p},t\}$-space to $\{\bvec{p}_0\}$-space, which we can use to perform a change of variables and obtain the plasma MDF as a function of time,
\begin{equation}\label{eq:MDF_solution_drift}
f ( \bvec{p},t ) = f_0 \big( \bvec{p}_0(\bvec{p},t) \big) \bigg( \frac{\gamma' \vphantom{p}_{\!\!\perp 0} \big( \bvec{p},t \big)}{\gamma' \vphantom{p}_{\!\!\perp}} \frac{p' \vphantom{p}_{\!\!\perp 0} \big( \bvec{p},t \big)}{p' \vphantom{p}_{\!\!\perp}} \bigg)^2 .
\end{equation}
This is analogous to using the method of characteristics to solve the Vlasov equation \eqref{eq:vlasov_drift}, since the momentum trajectories in equations \eqref{eq:approximate_motion_solution} are the characteristic curves in our specific case.

\begin{figure}
\centering
\includegraphics[width=\linewidth]{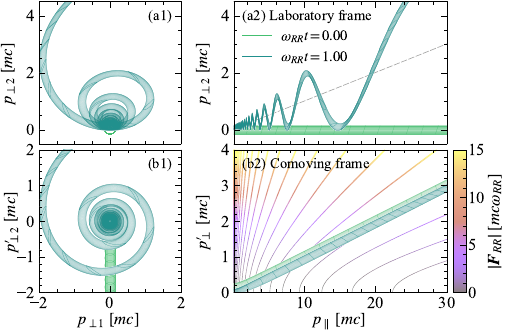}
\caption{
Deformation of an initially gyrotropic momentum space volume (green) into a spiral structure (blue) due to the presence of a drift velocity and the combined action of the Lorentz and radiation reaction forces.
The magnetic field points along $\bvec{e}_\parallel = \bvec{B}/B$ and the drift velocity along $\bvec{e}_{\perp 2}$.
In the laboratory frame (a), the guiding center of each individual particle shifts to the point $\gamma m \bvec{v}_d$, which lies near the dashed line in plot (a2), implying that all particles should gyrate around this curve.
Moreover, since particles gyrate with angular frequency $\omega_c/\gamma'$, particles with higher $p_\parallel$ gyrate at slower rates than particles with lower $p_\parallel$.
Over time, this builds up a phase difference between particles that deforms the phase-space volume into the spiral shape observed in plot (a1).
In the frame comoving with the drift velocity (b), the spiral is now centered around $p'_\perp = 0$, as shown in plot (b1).
The compression of phase-space volume due to radiative cooling now happens in $\{ p_\parallel,p'_\perp \}$-space, shown in plot (b2), where we have depicted the streamlines of the trajectories of particles in the background, with the color representing the magnitude of the radiation reaction force.
Similar to the case of a uniform B-field, the radiative cooling rate is still anisotropic and a non-linear function of particle momentum, but now it increases with the Lorentz-boosted perpendicular momentum, $p'_\perp$.
}
\label{fig:phasespace-third}
\end{figure}

We note that equation \eqref{eq:vlasov_drift} is obtained from Lorentz-boosting equation (16) in \cite{pablorings2024} to the frame comoving with the drift velocity, the De Hoffmann-Teller frame \citep{Hoffman-Teller}, where the plasma particles experience only the effect of a magnetic field.
Our main approximation relies on the fact that $|\bvec{v}_d| \ll c$, such that the time measured between the laboratory and comoving frames is essentially the same, $t' \simeq t$.
As such, the radiation reaction force terms (proportional to $\omega_{RR}$) assume the same form as for the case of a uniform B-field, but written in terms of Lorentz-boosted quantities, $p_\perp, \gamma_\perp \rightarrow p'_\perp, \gamma'_\perp$, suggesting that the compression of phase-space volume due to radiative cooling should now occur in $\{p_\parallel,p' \vphantom{p}_{\!\!\perp}\}$-space, which is depicted in plot (b2) of Fig.\ \ref{fig:phasespace-third}.
Likewise, the Lorentz force terms (proportional to $\omega_c$) are also written in terms of boosted quantities, $p_{\perp 2} \rightarrow p' \vphantom{p}_{\!\!\perp 2}$, meaning we can no longer neglect this term as is done in the original article.

When adding a drift velocity, the guiding center of the particles are shifted to the point $\gamma m \bvec{v}_d$, such that the whole MDF will also be dragged along in momentum space towards the direction of this drift velocity, which effectively breaks gyrotropy, that is, the axisymmetry of the system around the magnetic field lines, and guarantees that $\bvec{F}_L \cdot \boldsymbol{\nabla}_{\bvec{p}} f \neq 0$ for initially gyrotropic MDFs.
In addition to the usual phase-space compression due to the dissipative nature of radiation reaction, we thus also need to consider how the Lorentz force will deform a volume in momentum space.

The main contribution from the Lorentz force can be understood by recalling equation \eqref{eq:gyrophase} which, for $\omega_{RR} t \ll 1$, simplifies to
\begin{equation}\label{eq:gyrophase_expansion}
\psi = \frac{\omega_c}{\gamma' \vphantom{p}_{\!0}} t \left[ 1 + \frac{1}{2} \left( \frac{p' \vphantom{p}_{\!\!\perp 0}}{m c} \right)^2 \frac{\omega_{RR}}{\gamma' \vphantom{p}_{\!0}} t + \mathcal{O} \Big( \big( \omega_{RR} t \big)^2 \Big) \right] .
\end{equation}
According to the above relation, the particles gyrate in the plane perpendicular to $\bvec{B}$ with angular frequencies given initially by $\omega_c/\gamma' \vphantom{p}_{\!0}$, meaning that lower energetic particles will perform a larger number of rotations in momentum space than higher energetic ones.
Moreover, the gyration frequency of the particles will increase as their kinetic energies decay from radiation emission.
Over time, the relative difference in gyration frequencies between different regions of momentum space will deform the plasma MDF, even if it is initially gyrotropic, which then favors the formation of the spiral-shaped structures depicted in plots (a1) and (b1) of Fig.\ \ref{fig:phasespace-third}.
\\

We have assumed $\bvec{v}_d$ to be constant and equal for all particles, a premise that is only valid if the drift in question is the $\bvec{E} \times \bvec{B}$ drift, $\bvec{v}_E = \bvec{E} \times \bvec{B} / B^2$.
The curvature drift, for instance, should depend on the inertia of the particle, $v_c \propto \gamma v_\parallel \vphantom{v}^{\!2}$, and so the shift the plasma MDF suffers in momentum space will increase for particles with higher $p_\parallel$.
It can also be shown \citep{franciscothesis,franciscopaper} that particle drifts that depend on the momentum of the particle should decay as the particle radiates away its energy, which might suppress the effects of drift velocities in some scenarios.
Nevertheless, the intuition built here provides relevant insights on some of the numerical simulations presented later in this work (see bottom row of Fig.\ \ref{fig:KSvsSCHmdfreal} in Section \ref{subsec:RRCrealparams}, for instance).

\subsection{Ring distributions in dipolar magnetic fields}
\label{subsec:ringsnonideal}

Ring momentum distributions are characterized by their radius \citep{pablorings2024}, which determines where in phase space the bunching occurs. The ring radius $p_R(t)$ is defined as the value of the perpendicular momentum such that the perpendicular MDF, $f_\perp(p_\perp,t) = \int_{-\infty}^{+\infty} f(p_\parallel,p_\perp,t) dp_\parallel$, is maximum at every instant $t$, i.e., $\frac{\partial f_\perp}{\partial p_\perp}\bigr\rvert_{p_R} = 0$. Since spiral momentum distributions are no longer gyrotropic, their radius is no longer a well-defined property; thus, we generalize the properties of radiatively cooled plasma momentum distributions, previously studied in a uniform magnetic field, to more realistic cases.

\subsubsection{Maxwell-Jüttner distribution}
\label{subsubsec:MJrings}

The Maxwell-Jüttner distribution \citep{juttner1911} consists of the relativistic generalization of the Maxwell–Boltzmann distribution, and describes relativistic plasmas in thermal equilibrium. Following the technique detailed in \cite{pablorings2024} and approximating $f(p_\parallel,p_\perp,t) \approx \delta(p_\parallel)f_\perp(p_\perp,t) $, it is possible to show that the ring radius for an initial Maxwell-Jüttner distribution is, in the global laboratory frame:

\begin{equation}
    \label{eq:pRMJ}
    p_R(t) = m c \left[\frac{\gamma_b}{\omega_{RR}t} - \left(\frac{\gamma_bmc}{2 p_{th}\omega_{RR}t}\right)^2 \right] \ ,
\end{equation}

\noindent
where $\gamma_b$ and $p_{th} = \sqrt{m k_B T}$ are respectively the bulk Lorentz factor and the thermal momentum of the distribution. 

In the presence of drift velocities that significantly modify the plasma MDF, equation (\ref{eq:pRMJ}) is no longer exact, and instead we define an analogous quantity by averaging the MDF over the angle $\varphi = \arctan (p_{\perp 2}/ p_{\perp 1})$, before taking the value of $p_\perp$ at which the MDF reaches its maximum value:
\begin{equation}\label{eq:ring_radius_general_formula}
p_R(t) \equiv \underset{p_\perp}{\textrm{argmax}} \left( \frac{1}{p_\perp} \frac{1}{2 \pi} \int_0^{2 \pi} \!\!\!\! \int_{- \infty}^{+ \infty} \!\! f ( p_\parallel, p_\perp, \varphi, t) d p_\parallel d\varphi \right) ,
\end{equation}

\noindent
where the distribution function is written in cylindrical momentum coordinates $p_{\perp 1} = p_\perp \cos \varphi$ and $p_{\perp 2} = p_\perp \sin \varphi$, such that it transforms as $f ( p_\parallel, p_\perp, \varphi, t) = p_\perp f ( p_\parallel, p_{\perp 1}, p_{\perp 2}, t)$.

\medskip

The ring formation time is then defined as the time such that the ring radius $p_R$ reaches its maximum value. For the Maxwell-Jüttner distribution, in the global laboratory (or fiducial) frame, this quantity is determined by:
\begin{equation}
    \label{eq:tRMJ}
    t_R = \frac{\gamma_b m^2c^2}{2\omega_{RR} p_{th}^2}\ ,
\end{equation}

\noindent
which differs from its Maxwell-Boltzmann counterpart only by a factor of $mc/p_{th}$. This is due to the fact that, for the same $p_{th}$, the Maxwell-Jüttner distribution possesses more energetic particles than the Maxwell-Boltzmann, on average, causing the ring to form faster. 
Even though equation (\ref{eq:tRMJ}) was derived for ring-shaped distributions in uniform magnetic fields, we note that the spiral-shaped MDFs studied in Section \ref{subsec:spirals} still develop inverted Landau populations, $\partial f / \partial p_\perp > 0$, in timescales of the same order of magnitude of $t_R$ \citep{franciscothesis}, and we therefore still use it to identify the relevant timescale of our simulations.

The radiation reaction cooling process of a plasma can therefore be characterized from this quantity. For $t < t_R$, the bunching region of the momentum distribution function expands outwards in momentum space, and the ring radius achieves its maximum value at $t=t_R$, while the distribution keeps contracting as a whole. 
From then on, the ring radius starts to decrease, leading to an accelerated increase of $\partial f / \partial p_\perp$. At later times, $t > t_R$, the plasma has radiated most of its energy, and its momentum distribution has a very high gradient, thereby fulfilling the necessary conditions for the efficient emission of coherent radiation via the electron cyclotron maser instability \citep{pabloECMIrings}.

A plasma with a Maxwell-Jüttner momentum distribution develops inverted Landau populations as long as $p_{th} > mc/3$ \citep{pablorings2024}, implying a minimum temperature of $T \ge 7 \times 10^8 \unid{K}$. This is also the criterion that determines when the ECMI-powered emission process stops: the condition $p_R(t) = m c/3$ defines the emission time $t_{em}$, which expresses how long the emission process is sustained. When $t_{em}$ is smaller than the timescale for the maser onset $t_o$, radiation is still emitted as a single burst, instead of continued emission \citep{pabloECMIrings}. In the global frame, these timescales are formulated as:
\begin{equation}
    \label{eq:totem}
    t_o = \frac{\gamma_b}{\omega_{p}} \sqrt{\frac{mcB_{sc}}{\sqrt{3}\pi\alpha_S p_{th} B}} \ \ , \ \ t_{em} = \frac{4}{3\sqrt{3}}\frac{\gamma_b}{\omega_{RR}} \ .
\end{equation}

\noindent
where $\omega_p = \sqrt{4\pi nq^2/m}$ is the plasma frequency, and constant fields and a flat metric have been assumed. We will use these scalings to inform the estimates and guide the discussion for the nonideal scenarios.

\subsubsection{Dipolar magnetic field}
\label{subsubsec:dipolarrings}

The global electromagnetic field configuration of a neutron star can be considered, at zero order, to consist of a magnetic dipole field, characterized by a magnetic moment $\mu$ and displacement angle $\chi$ between the dipole moment and the angular velocity \citep{deutschfields,rezzollafields,ruiBFDcorrection}. For simplicity, in this work, we will always consider $\chi=0$. In a dipolar field, the amplitude of the magnetic field is no longer constant and independent of the spatial position. 
It is thus expected that the initial position of the plasma relative to the dipole affects the radiative cooling process and, consequently, the resulting momentum distribution function. 

These effects can be quantitatively assessed by deriving the ring radius for an initial Maxwell-Jüttner plasma in a dipolar magnetic field. Approximating the plasma as propagating along the magnetic field with velocity $v \approx c$ and neglecting the contribution of the second term in equation (\ref{eq:pRMJ}), which is negligible, we find that the ring radius in the global frame is given by $p_R(r) \approx \frac{3\gamma_b m^2 c^3 B_{sc}}{2 \alpha_S q} \left( \int_{r_i}^{r} B^2(r') dr' \right)^{-1}$, where $r_i$ is the initial position of the plasma. For a dipolar field $B(r) \approx B_* (R_*/r)^3$, the integral in the previous expression yields:

\begin{equation}
    \label{eq:pRdipole}
    p_R(r) = \frac{5\gamma_b m c^2}{\omega_{RR}^*R_*^6} \frac{1}{1/r_i^5 - 1/r^5} \ .
\end{equation}

\noindent
where $\omega_{RR}^*$ is the radiation damping frequency for the magnetic field intensity at the magnetic pole $B_*$. In neutron stars, field strengths range from $B_* \sim 10^8-10^{10} \unid{G}$ in millisecond pulsars, to $B_* \sim 10^{11}-10^{13} \unid{G}$ in radio pulsars and $B_* \sim 10^{14}-10^{15} \unid{G}$ in magnetars.

For sufficiently large distances from the dipole, the magnetic field diminishes to such an extent that the MDF may never develop an effective positive gradient, due to the inefficiency of the cooling process. This behaviour is more apparent from the dipole ring formation time, which can be derived from the Vlasov equation as \citep{franciscothesis}:
\begin{equation}
    \label{eq:tRdipole}
    t_R^D = \left[ \frac{R_*}{c}\left(\frac{R_*}{r_i} \right)^5 - 5 t_R \right]^{-\frac{1}{5}} - \frac{r_i}{c} \ .
\end{equation}

\noindent
where $t_R$ is the ring formation time for a uniform magnetic field, expressed by equation (\ref{eq:tRMJ}). It is worth noting that equation (\ref{eq:tRdipole}) also applies for converting any timescale derived for a uniform magnetic field to its dipolar field counterpart. 

Equation (\ref{eq:tRdipole}) confirms that, for any given compact object, there is a finite initial distance above which, if a pair plasma beam is injected, it cannot effectively develop inverted Landau populations. A lower bound for this quantity is given by the injection radius such that $t_R \to \infty$:
\begin{equation}
    \label{eq:rimaxdipole}
    r_i < r_i^\text{max} = R_* \left( \frac{5\gamma_b m^2c^3}{2 R_* \omega_{RR}^* p_{th}^2} \right)^{-\frac{1}{5}} \ .
\end{equation}

\noindent
Equation (\ref{eq:rimaxdipole}) assumes that the plasma parameters $\gamma_b$ and $p_{th}$ are constant along the trajectory; in general, we expect deviations to be dominated by the decrease in temperature due to the cooling process. 
Therefore, we stress that this result consists of a minimum lower bound, and that plasmas injected at distances slightly above $r_i=R_* \left( 5c t_R/R_* \right)^{-1/5}$ may still develop inverted ring momentum distributions. However, their ring radius will never reach its maximum value. 

\smallskip

For the efficient emission of coherent radiation, the plasma must be injected sufficiently close to the star such that radiative cooling is strong enough to trigger the development of population inversion, and, at the same time, sufficiently far from the star such that the ring does not collapse to the lowest Landau level before the onset of the ECMI. Since the first quantum Landau momentum level is given by $p_L = \frac{B}{B_{sc}} mc$, the time needed for the ring distribution to become fully degenerate is, in the global frame, in the order of:

\begin{equation}
    \label{eq:tLandau}
    t_L = \frac{4 \gamma_b B_{sc}}{9 \omega_{RR} B} \ ,
\end{equation}

\noindent
where we have assumed a uniform magnetic field. 

Therefore, the plasma must not only satisfy $t_o < t_L$, but also $p_R > p_L$, at all times. For a Maxwell-Jüttner plasma in a dipolar magnetic field, the function $p_R/p_L$, as derived from equation (\ref{eq:pRdipole}), has a minimum at $r = \sqrt[^5]{8/3} \ r_i$. This implies that a higher bound for the minimum injection distance (to guarantee that the ring distribution never becomes degenerate) can be obtained from equation (\ref{eq:pRdipole}), by requiring that the minimum of $p_R/p_L$ exceeds unity. This condition is written as:
\begin{equation}
    \label{eq:pRpL}
    r_i > r_i^\text{min} = R_* \left( \frac{3}{32} \sqrt[^5]{\frac{2}{9}} \frac{\omega_{RR}^* B_* R_*}{\gamma_b c B_{sc}} \right)^\frac{1}{8} \ .
\end{equation}

These results imply that synchrotron-induced ECMI coherent emission will always occur, at least as a single outburst of radiation, as long as conditions (\ref{eq:rimaxdipole}) and (\ref{eq:pRpL}) are simultaneously satisfied. 
Exemplifying numerical estimates for these criteria can be obtained by considering the scaling engineering formulas for $r_i^\text{min}$ and $r_i^\text{max}$, which yield:
\begin{align}
    \label{eq:riminmaxscalings}
    r_i^\text{min} [42.3 \unid{km}] &= \frac{R_*^{9/8} [10 \unid{km}] \, B_*^{3/8} [10^{12} \unid{G}]}{\gamma_b^{1/8} [10^3]} \ ; \\ \nonumber
    r_i^\text{max} [1916.9 \unid{km}] &= \frac{R_*^{4/5} [10 \unid{km}] \, B_*^{2/5} [10^{12} \unid{G}] \, p_{th}^{2/5} [100 mc]}{\gamma_b^{1/5} [10^3]} \ .
\end{align}

\noindent
These distances encompass a wide range of parameters, and are thus generally compatible with the regions where pair plasmas are produced in acceleration gaps throughout the magnetosphere \citep{bransgrove2023radio}.

\subsection{Astrophysical coherent emission from synchrotron ECMI}
\label{subsec:astrophysicalECMI}

The conditions required for synchrotron-powered coherent emission from the ECMI to occur can arise generically during the propagation of a radiatively cooled relativistic plasma through the magnetosphere as long as a momentum distribution with an inverted Landau population is present, as explored in the previous section. The instability becomes efficient in regions where the local plasma frequency is slightly below the cyclotron frequency, $\omega_{p} \lesssim \omega_{c}$, marking the point where emission takes place. 
In the beam frame, the resulting radiation is a highly elliptically polarized X-mode, which is very close to linear polarization \citep{pabloECMIrings}. After Lorentz boosting, the radiation propagates nearly parallel to the magnetic field, appearing in the observer frame as circularly polarized and aligned with the magnetic field direction \citep{alexandrov1984principles},  consistent with the emission properties of several observed fast radio bursts (FRBs) and pulsars \citep{coherenttemperature1,lorimer2005handbook,frbreview2,frborigincompactobj2,FRBalign}. 

To illustrate the working of this mechanism, let us consider a concrete example: a pulsar of radius $R_* = 10^6 \unid{cm}$ and polar surface field $B_* = 5\times10^{10} \unid{G}$ with a rotation period $P = 2\pi/\Omega_* = 0.1 \unid{s}$, corresponding to a light-cylinder radius $R_{LC} = c/\Omega_* =  477\,R_*$. The plasma beam is a Maxwell-Jüttner pair plasma injected at $r_i=1.3\,R_*$ with $\gamma_b=2000$ and $p_{th}=25 \, mc$, with a density profile $n = n_* (R_*/r)^3$, where $n_* = 7 \times 10^{13} \unid{cm^{-3}}$ is the Goldreich–Julian density at the stellar surface assuming a pair multiplicity $\kappa = 10^3$ \citep{goldreichjulian}. These parameters are representative of physically realistic systems, aligning with typical estimates of the magnetic fields $B_* \sim 10^{10}-10^{13} \unid{G}$ and rotation periods $P \sim 0.1-10\,\unid{s}$ of radio pulsars \citep{pulsarcatalog2,pulsarcatalog}, and with the Lorentz factor $\gamma \sim 10^2-10^4$ and pair multiplicity $\kappa \sim 10^2-10^5$ of the secondary pairs produced in pair cascades \citep{qedmagnetosphere1,pulsarmagconfirm1,philippov2018pairs,pulsarmagemission}.

For these parameters, the relativistic electron cyclotron frequency $\omega_{c}/\gamma$ matches the radio band $\nu = 10\,\unid{GHz}$ in the observer frame when the magnetic field has decreased to $B \approx 3.6 \times 10^3 \unid{G}$, corresponding to a radial distance of $r \approx 0.5\,R_{LC}$. At this location, the electron plasma frequency is $\omega_{p} \approx 1.26 \unid{GHz}$, which fulfills exactly the condition for the efficient emission of coherent radiation via the ECMI. 
For an inverted ring distribution to exist in the emission region, equations (\ref{eq:rimaxdipole}) and (\ref{eq:pRpL}) restrict the initial injection distance to the range $1.2 < r_i/R_* < 28$, which is concurrent with the previously chosen value $r_i = 1.3\,R_*$. This injection distance is in agreement with typical estimates of the spatial extent of the polar-cap gap, where the electron–positron pairs that dominate the emission of synchrotron radiation are produced in pulsar magnetospheres \citep{ruderman1975pair,arons1979pair,muslimov1992general,timokhin2010paircascade,qedmagnetosphere1}. 

The synchrotron-cooling-induced ECMI accounts for the brightness temperatures and spectral features associated with these coherent astrophysical phenomena \citep{pablothesis}, and is thus capable of generating extreme brightness temperatures on the order of $10^{30} \unid{K}$ without fine-tuned conditions \citep{frborigincompactobj1}.
Altogether, these considerations establish that the ECMI driven by radiatively cooled inverted momentum distributions provides a self-consistent, first-principles explanation of coherent emission in neutron star magnetospheres based solely on kinetic processes, and motivate a closer examination of how such conditions can arise and evolve in realistic astrophysical configurations.

\section{Numerical study of radiatively cooled plasma dynamics in curved spacetime}
\label{sec:4} 

\subsection{Simulation framework}
\label{sec:simulationframework}

The dynamics of radiatively cooled plasmas in neutron star magnetospheres are usually studied numerically with general-relativistic particle-in-cell (GR-PIC) codes \citep{grpic1,grpic2,grpic3,osirisgr}. 
For the purposes of studying how curved spacetime effects modify the formation and evolution of inverted momentum distributions, it suffices to assume that the electromagnetic fields are stationary in time and determined by the global structure for an aligned rotating magnetic dipole in curved spacetime \citep{rezzollafields}, since the timescale of evolution of the fields is typically much longer than the characteristic timescales for particle motion and synchrotron emission in the inner magnetosphere \citep{pulsarmagconfirm5,luminosityformula2,statfieldsapprox4,fieldevolution}. 

Therefore, the stationary field approximation is valid as long as the timescales associated with the dynamics of inverted distributions, i.e., the ring formation time, are much smaller than the dynamical evolution of collective plasma effects. In addition, we note that the electromagnetic field configuration may be modified by gravitational charge-dependent drifts. Such corrections are expected to be smaller than the GR-induced modifications already incorporated into the external field expressions, and will only alter the plasma distribution after the onset of collective effects. Nevertheless, we conjecture that the instabilities developed from the population inversion observed here will not be significantly different.

\medskip

In our physical system, and before the onset of collective plasma processes, the hierarchy of physical timescales is described by:
\begin{equation}
    \label{eq:scalehierarchy}
    \frac{2\pi}{\omega_c} \ll t_{R} \lesssim t_c \ll t_o \ll \frac{R_*}{c} \ll P \ ,
\end{equation}

\noindent
where $t_c = (\omega_{RR} \gamma \sin^2{\theta_0})^{-1}$ is the synchrotron cooling time for one particle, with ${\theta_0}$ the angle between the particle trajectory and the local magnetic field line \citep{rcoolingtime}, $t_R$ is the ring formation time in the global fiducial frame (\ref{eq:tRMJ}), and $P$ is the stellar rotation period. The assumption of field stationarity thus applies until the onset of the kinetic instabilities triggered by radiative cooling, $t_o$. 

In order to understand the dynamics of synchrotron-cooled plasmas in arbitrary spacetime configurations, we follow an ensemble of test particles whose dynamics are advanced with a parallelized particle pusher that integrates the equations of motion (\ref{eq:3+1eqmom}) for every particle in the ensemble. The particle pusher is described in detail in Appendix \ref{app:ppusher}. By propagating a sufficiently large number of particles, we reconstruct the evolution of the distribution function, thereby obtaining a detailed picture of the phase-space dynamics without requiring the full complexity and computational cost of a self-consistent evolution \citep{franciscothesis}. 

The simulation domain is axisymmetric because the magnetic field configuration is an aligned rotating dipole, characterized by its magnetic moment $\mu$. When comparing different spacetime metrics, we keep the stellar magnetic moment fixed, to preserve the star properties; the corresponding intensity at the magnetic pole is given by $B_* = 2\mu/R_*^3$, in flat spacetime. Particles are initialized in a spatially uniform, small localized volume at a given position $(r_i,\theta_i)$, in the polar cap, $\theta < \arcsin{\sqrt{R_*/R_{LC}}}$. The plasma is launched outwards along the magnetic field lines, with initial momenta sampled from a boosted Maxwell-Jüttner distribution $f(\bvec{p}) \propto \exp{\left(-\gamma_b(\gamma mc^2-\bvec{v}_b \cdot \bvec{p})/p_{th}^2\right)}$ \citep{boostedMJ2,boostedMJ1}. 

\subsection{Realistic parameter simulations}
\label{subsec:RRCrealparams}

\begin{figure}
    \centering
    \includegraphics[width=\linewidth]{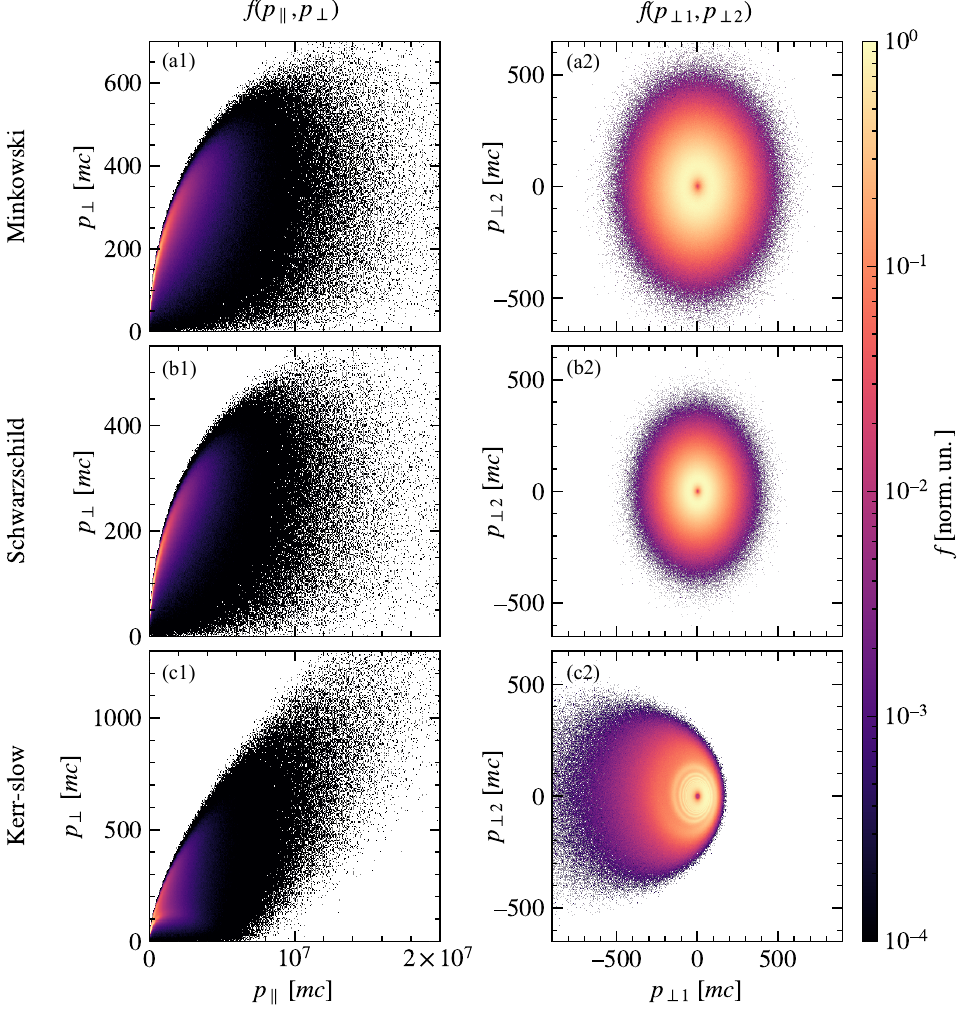}
    \caption{Momentum distribution function at $t = 10\,t_R$ for an ensemble of $10^7$ electrons initialized at $r_i=1.3\,R_*$ and $\theta_i = 5{}^\circ$ from a Maxwell-Jüttner distribution with $\gamma_b = 2000$ and $p_{th} = 25\,mc$, subject to a magnetic dipole field with flat-spacetime polar surface intensity $B_* = 5\times10^{10} \unid{G}$. In the top panel, particles were launched in non-rotating Minkowski spacetime, while in the middle panel the metric is Schwarzschild with stellar compactness $r_s/R_* = 0.5$, and in the bottom panel the metric is Kerr-slow with $r_s/R_* = 0.5$ and stellar rotation period $P = 0.1 \unid{s}$. Inverted spiral-shaped distributions are found to emerge in different curved spacetime configurations as well. Ring-shaped momentum distributions emerge in Schwarzschild spacetime, while in Kerr-slow spacetime the distributions are spiral-shaped, due to the $\bvec{E} \times \bvec{B}$ drift.}
    \label{fig:KSvsSCHmdfreal}
\end{figure}

Realistic numerical simulations of the magnetospheres of astrophysical compact objects pose the inherent problem of encompassing vastly different spatial and temporal scales. 
For this reason, we begin our numerical analysis with a small set of simulations employing parameter values representative of physical neutron star magnetospheres, whose aim is to characterize directly the radiation reaction cooling process under physically realistic conditions. In these simulations, whose computational cost is approximately 500k CPU core hours each, we defined the system parameters as those introduced in Section \ref{subsec:astrophysicalECMI}, to ensure consistency with theoretical estimates. 

Given the extremely short simulation times $t=10\,t_R \sim 10^{-7} R_*/c$ accessible in realistic-parameter simulations, particle trajectories remain confined to a very narrow spatial region. Since particles are initialized close to the rotation axis, at a small polar angle, the plasma effectively experiences a locally uniform radially directed magnetic field, with negligible variation in curvature. This allows for a direct comparison with the analytical theory developed in Section \ref{subsec:ringsnonideal}. 

\medskip

We first assess how the presence of curved spacetime affects the formation of inverted distributions by evolving an ensemble of $10^7$ electrons in a non-rotating magnetic dipole field configuration for Minkowski, Schwarzschild, and Kerr-slow background metrics. The momentum distribution functions $f(p_\parallel,p_\perp)$ at $t=10 \, t_R$ are portrayed in Fig. \ref{fig:KSvsSCHmdfreal}. The emergence of population inversion is immediately visible in Fig. \ref{fig:KSvsSCHmdfreal}: particles cluster in a much narrower region in the $p_\perp$ plane than along the $p_\parallel$ direction, confirming that the mechanism through which radiative cooling creates inverted momentum distributions in flat spacetime persists in curved spacetime. Moreover, particles possess lower parallel momenta in curved than in flat spacetime, due to the gravitational acceleration that is directed radially inwards.

\begin{figure}
    \centering
    \includegraphics[width=\linewidth]{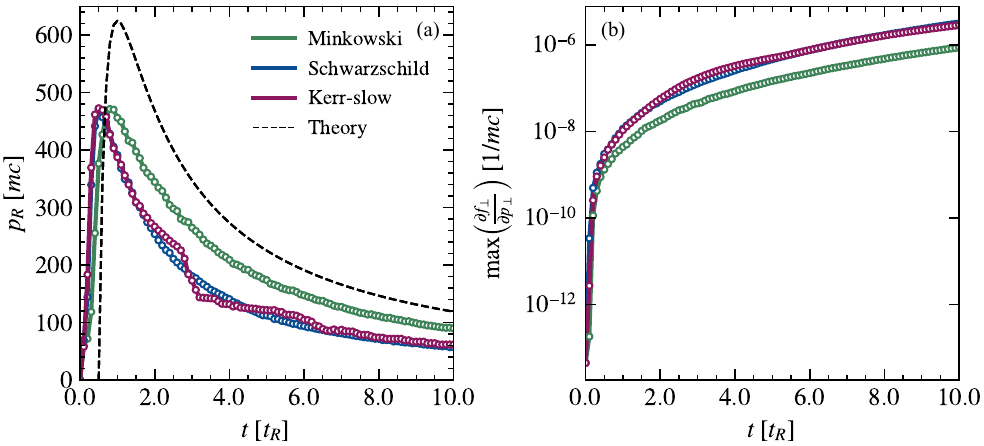}
    \caption{Evolution of the ring radius and maximum perpendicular momentum gradient for an ensemble of $10^7$ electrons initialized at $r_i=1.3\,R_*$ and $\theta_i = 5{}^\circ$ from a Maxwell-Jüttner distribution with $\gamma_b = 2000$ and $p_{th} = 25\,mc$, subject to a magnetic dipole field in non-rotating Minkowski, Schwarzschild and Kerr-slow spacetimes with flat-spacetime polar surface intensity $B_* = 5\times10^{10} \unid{G}$. For the Kerr-slow metric, the stellar rotation period is $P = 0.1 \unid{s}$, while for Schwarzschild and Kerr-slow the stellar compactness is $r_s/R_*=0.5$. The dashed line represents the theoretical prediction, equation (\ref{eq:pRMJ}), originally derived in the proper frame of the plasma. When the stellar mass is taken into account, the ring radius diminishes and the perpendicular momentum gradient increases, while the stellar rotation has little influence due to its reduced value.}
    \label{fig:realparamrradius}
\end{figure}

These momentum distribution functions possess well-defined shapes. In Schwarzschild spacetime, the only additional force acting on the particles is gravity, which is radially directed and thus has no effect in the transverse momentum space. Therefore, both the $\bvec{g} \times \bvec{B}$ and the curvature drifts are suppressed, and the MDF retains the ring-like shape characteristic of uniform fields in flat spacetime. In contrast, in Kerr-slow spacetime, the perpendicular momentum distribution develops a pronounced spiral-like structure that arises from the $\bvec{E} \times \bvec{B}$ drift created by the stellar electric field. As demonstrated in Section \ref{subsec:spirals}, this drift displaces the ring as a whole, imprinting an angular dependence that manifests as a spiral pattern. 
The evolutions of the ring radius and maximum perpendicular momentum gradient are presented in Fig. \ref{fig:realparamrradius}, for all the different simulated spacetime metrics.

The maximum perpendicular momentum gradient is always positive, indicating that radiation reaction sustains population inversion throughout the entire simulation. 
Furthermore, the ring radius in flat spacetime is larger than in curved spacetime, confirming that the gravitational acceleration constricts the momentum distribution towards smaller momenta, since it is opposed to the bulk motion of the beam. This behaviour can also be observed in Fig. \ref{fig:compactnessrradius}, in which a further comparison between the curved and flat spacetime cases is shown.

The ring radii agree qualitatively with the theoretical calculation (\ref{eq:pRMJ}), but their evolutions do not quantitatively follow the prediction. The analytic results are originally derived in the proper frame of the plasma, where the radiative cooling process can be treated locally at a fixed $p_\parallel$. Therefore, the boost of the distribution at a specific proper time $\tau$ to a global laboratory frame at high energies $\gamma_b \gg 1$ requires a more computationally demanding approach \citep{sfmartins2010}. Nevertheless, the resulting MDFs possess the same properties and follow a similar qualitative trend as the theoretical prediction, confirming the analytic theory developed in Sections \ref{subsec:spirals} and \ref{subsec:ringsnonideal}. 

Preliminary PIC simulations \citep{franciscopaper} show that spiral distributions develop momentum-space features that support an inverted Landau population, and will therefore remain unstable to the ECMI. However, we emphasize that these are novel phase-space structures distinct from standard ring distributions and with distinct radiation features, even though they still possess a positive perpendicular momentum gradient. These simulations confirm that inverted momentum distributions still arise in realistic astrophysical configurations, and highlight the need for a broader exploration of parameter space.

\subsection{Rescaled parameter simulations}
\label{subsec:rescaledparams}

As previously outlined, simulating the full dynamical evolution of the system is not computationally feasible due to the large scale separation. Therefore, we study the dependence of the radiative cooling process on the curved spacetime parameters, namely the stellar mass and rotation, resorting to rescaled parameter simulations. The physical correctness of this approach is maintained provided that the hierarchy (\ref{eq:scalehierarchy}) between the orders of magnitude of the relevant characteristic timescales is preserved \citep{magnetospherereview,paramrescalingPIC}. As such, the simulation parameters are defined such that $2\pi/\omega_c \sim 0.1\,t_R \sim 0.5\,t_c \sim 0.01\,R_*/c \sim 0.1\,P$ at the stellar surface, while the simulation timestep interval is chosen to always fully resolve the gyration orbits of particles. 

\subsubsection{Stellar compactness and Schwarzschild spacetime}
\label{subsec:RRCschwarzschild}

\begin{figure}
    \centering
    \includegraphics[width=\linewidth]{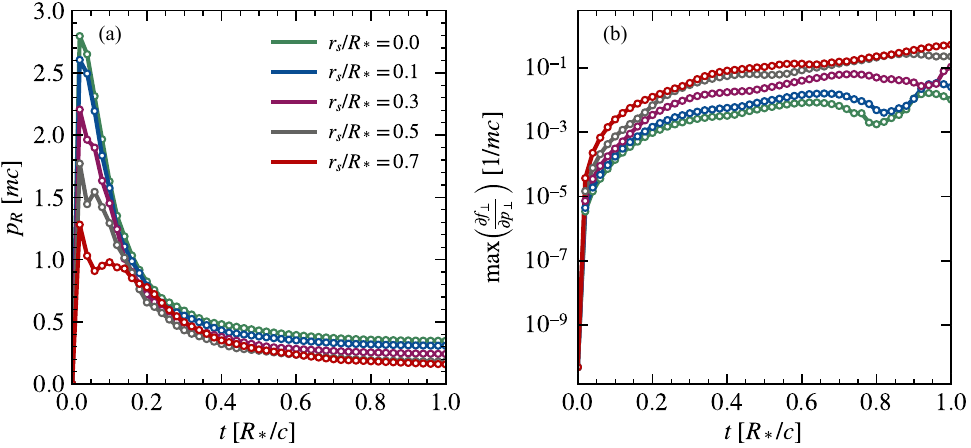}
    \caption{Evolution of the ring radius and maximum perpendicular momentum gradient for an ensemble of $10^7$ electrons initialized at $r_i=R_*$ and $\theta_i = 5{}^\circ$ from a Maxwell-Jüttner distribution with $\gamma_b = 100$ and $p_{th} = 10\,mc$, subject to a non-rotating magnetic dipole field in Schwarzschild spacetime with flat-spacetime polar surface intensity $B_*/B_{sc} = 1$. The simulations were performed for varying values of the stellar compactness $r_s/R_* = \{0, 0.1, \ 0.3, \ 0.5, \ 0.7\}$. Increasing the stellar compactness reduces the ring radius and augments the maximum perpendicular momentum gradient.}
    \label{fig:compactnessrradius}
\end{figure}

To build on the physical intuition developed by comparing flat and curved spacetime configurations in Section \ref{subsec:RRCrealparams}, we performed a series of simulations in which the stellar compactness is varied at zero angular velocity. The evolution of the ring radii and maximum perpendicular momentum gradient is displayed in Fig. \ref{fig:compactnessrradius}, for different values of the stellar compactness.

The simulation results displayed in Fig. \ref{fig:compactnessrradius} indicate that gravity diminishes the ring radius and increases the gradient $\frac{\partial f_\perp}{\partial p_\perp}$. This is partially due to the fact that, at fixed magnetic dipole moment, the amplitude of the electromagnetic fields of a rotating magnetic dipole in curved spacetime is higher than in flat spacetime: gravity compresses the field lines, enhancing the flux densities near the stellar surface \citep{bfieldstrongerGR2,bfieldstrongerGR1,muslimov1992general,ruiBFDcorrection}. Furthermore, we verified that for simulations employing the same polar surface field in all spacetime configurations, instead of the same magnetic moment, the increase in perpendicular gradient is smaller, but still occurs at later stages of the ring formation, due to the gravitational acceleration.

Gravity is radially directed and thus globally diminishes the parallel momentum of the particle ensemble without affecting its perpendicular component, leading to a decrease in the Lorentz factor. This causes the cooling rate to augment, thereby accelerating the cooling process. We note that this effect is a consequence of the geometry of the problem due to gravity being opposed to the outgoing bulk motion of the plasma, as hinted in Section \ref{subsec:RRCrealparams}; its intensity increases with the decrease in polar angle, as the magnetic field becomes aligned with the radial direction. This is an additional source for the observed increment of the perpendicular momentum gradient in Schwarzschild spacetime. 

\subsubsection{Stellar rotation and Kerr-slow spacetime}
\label{subsec:RRCKS}

We now extend the analysis to rotating spacetimes, including the stellar rotation. This causes inertial frames to co-rotate with the central object, giving rise to a non-zero azimuthal velocity component through frame-dragging, and induces a stellar quadrupolar electric field that alters the trajectories of the particles. To explore these effects, we simulated an ensemble of $10^7$ electrons in a rotating magnetic dipole field configuration at fixed compactness $r_s/R_*=0.5$, for different values of the stellar angular velocity $\Omega_*$. 

\begin{figure}
    \centering
    \includegraphics[width=\linewidth]{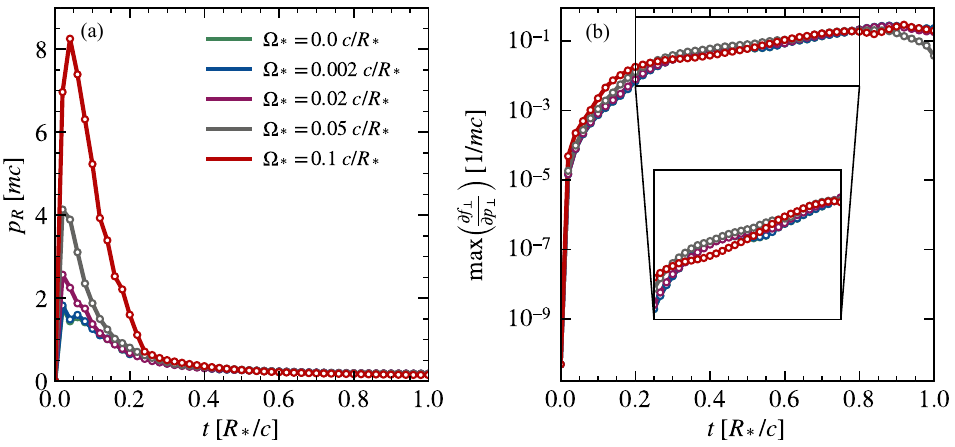}
    \caption{Evolution of the ring radius and maximum perpendicular momentum gradient for an ensemble of $10^7$ electrons initialized at $r_i=R_*$ and $\theta_i = 5{}^\circ$ from a Maxwell-Jüttner distribution with $\gamma_b = 100$ and $p_{th} = 10\,mc$, subject to a rotating magnetic dipole field in Kerr-slow spacetime with flat-spacetime polar surface intensity $B_*/B_{sc} = 1$ and stellar compactness $r_s/R_* = 0.5$. The simulations were performed for varying values of the angular velocity $\Omega_* = \{0, 0.002, \ 0.02, \ 0.05, \ 0.1\} \, c/R_*$. The ring radius increases with the stellar rotation due to the broadening of the distribution in $p_\perp$ caused by the frame-dragging effect, overcoming the effect of the gravitational acceleration, while the maximum perpendicular momentum gradient shows no dependence on the angular velocity at fixed compactness.}
    \label{fig:fdrradius}
\end{figure}

The simulation results in Fig. \ref{fig:fdrradius}a) reveal that the ring radius increases significantly with the stellar rotation. This is a direct consequence of the frame-dragging effect, which is azimuthally directed and therefore always perpendicular to the magnetic field in the aligned dipole configuration, causing the distribution to broaden along the $p_\perp$ direction in Kerr-slow spacetime. Since the maximum perpendicular momentum gradient is independent of the angular velocity at fixed compactness, as demonstrated in Fig. \ref{fig:fdrradius}b), population inversion will be preserved across longer timescales as compared with other non-rotating spacetimes, in which a constant source of perpendicular momentum does not exist. We note that this effect is especially noticeable in longer evolutions of the plasma, for $t > R_*/c$, and at higher values of angular velocity, beyond the slow-rotation condition. 

We also note that the radiative cooling timescales in curved spacetime will always be longer than in flat spacetime due to gravitational time dilation, irrespective of whether the metric is rotating. The temporal evolution of the distribution function is governed by the general-relativistic Vlasov equation (\ref{eq:3+1radvlasov}), which, for spatially homogeneous plasmas, can be generically recast as:
\begin{align}
    \label{eq:vlasovtimedilation}
    &\frac{1}{\alpha}\frac{\partial f}{\partial t} + \bvec{\nabla_p} \cdot \left(\frac{
f}{\alpha} \frac{d\bvec{p}}{dt}\right) = 0 \ . 
\end{align}

The global time coordinate $t$ thus becomes rescaled by the lapse function, which is always less than unity in the spacetimes we consider. As a consequence, all processes that are local in proper time become further dilated when measured in the global time coordinate. This can also be derived from the relation between proper and coordinate time, which in curved spacetime reads $\frac{d \tau}{dt} = \frac{\alpha}{\gamma}$. 

\medskip

The phase-space evolution observed in the rescaled simulations mirrors that obtained with realistic parameters. The perpendicular momentum gradient increases in curved spacetime metrics due to the effect of the gravitational acceleration, which also causes the ring radii to be smaller than in Minkowski spacetime (see Section \ref{subsec:RRCschwarzschild}). On the other hand, both the frame-dragging effect and the rotation-induced electric field are weak due to the small values of the angular velocity, and therefore do not have a pronounced impact on the ring radius and on the distribution gradient. This agreement provides direct validation of the numerical rescaling procedure adopted in this work, and strengthens the assumption that the extrapolated reduced system faithfully captures the underlying physics.

Overall, our simulations demonstrate that the perpendicular momentum gradient of the distribution function increases with the stellar mass and is not altered significantly by the stellar rotation. Since the growth rate of the ECMI is proportional to $\frac{\partial f_\perp}{\partial p_\perp}$, \citep{ecmitheory1,ecmitheory2,pabloECMIrings}, we conclude that curved spacetime effects enhance the conditions for the onset and amplification of coherent radiation emission. This is a strong indication that gravity may reinforce this coherent emission mechanism.

\section{Conclusions}
\label{sec:conclusions}

In this work, we have presented a systematic investigation of radiation reaction cooling in realistic astrophysical conditions. We extended the underlying theoretical formalism to include general relativity, and established the minimum and maximum injection distances required for inverted momentum distributions to always sustain population inversion in the non-ideal configurations characteristic of neutron star magnetospheres, by taking into account the timescales for radiative cooling to the lowest Landau level.

Our results confirm that radiation reaction cooling triggers the formation of anisotropic momentum distributions with inverted Landau populations under non-uniform field geometries and in curved spacetime. We demonstrated analytically that in the presence of any generic drift velocity, radiation reaction drives the formation of inverted spiral-shaped momentum distributions: this behaviour can be caused by electric fields, field inhomogeneities, or even external forces of non-electromagnetic nature. A detailed investigation of the stability and radiation properties of spiral momentum distributions, accounting for collective plasma effects, is already the subject of ongoing work \citep{franciscopaper}.


Furthermore, we showed, resorting to simulations, that accounting for the stellar compactness amplifies the gradient of the perpendicular MDF, thereby increasing the growth rate of the ECMI compared to flat spacetime. Curved-spacetime effects were also found to prolong the persistence of the inverted spiral momentum structure, both due to gravitational time dilation and to the frame-dragging effect, which acts as a source of perpendicular momentum. This analysis confirms that realistic astrophysical environments preserve, and may also improve the conditions necessary for radiative-cooling-powered emission of coherent radiation in the magnetospheres of neutron stars.

\section*{Acknowledgements}

We would like to thank A. Bransgrove and T. Silva for discussions. This work was partially supported by the Foundation for Science and Technology, I.P. (FCT) through the project X-Maser (No. 2022.02230.PTDC). Simulations were performed at MareNostrum 5 (Barcelona Supercomputing Center, Spain), funded by the FCT projects \textit{Realistic simulations of relativistic plasmas in astrophysical and laboratory environments} (No. 2024.07895.CPCA.A3, https://doi.org/10.54499/2024.07895.CPCA.A3), and \textit{Fireball Beams and Matter Antimatter Plasmas at CERN} (No. 2025.00289.CPCA.A3). IPFN's activities were supported by FCT through the project UID/50010/2025 (https://doi.org/10.54499/UID/50010/2025).

\section*{Data Availability}

All data needed to evaluate the conclusions are present in the paper. The simulation results presented in this work are available online via Zenodo \citep{zenodo}.


\bibliographystyle{mnras}
\bibliography{Thesis}


\appendix

\section{Radiation reaction in curved spacetime}
\label{app:RRG}

Classical radiation reaction in curved spacetime is formally described by the DeWitt-Brehme equation \citep{RRdewittbrehme1,RRdewittbrehme2}, which generalizes the classical Lorentz-Abraham-Dirac force \citep{diracRR} to account for radiative effects that occur without the presence of external electromagnetic fields, due to the equivalence principle. As in flat spacetime, the general-relativistic Landau-Lifshitz force is obtained from the DeWitt-Brehme equation by reduction of order, and is formulated as \citep{RRGRLL}:
\begin{align}
    \label{eq:dewittbrehmeRR}
    F_{RR}^\mu  &= \frac{2q^3}{3m} \left[ (\nabla_\alpha F^{\mu}_{\nu}) u^\nu u^\alpha + \frac{q}{m} ( F^{\mu\nu} F_{\nu\alpha} u^\alpha + F^{\alpha\beta} F_{\beta\nu} u_\alpha u^\nu u^\mu) \right] \nonumber \\ &+ \frac{q^2}{3m} (R^\mu_\nu u^\nu + R_{\nu\alpha} u^\nu u^\alpha u^\mu) + \frac{2q^2}{m} f^{\mu\nu}_{\text{tail}} u_\nu \ ,
\end{align}

\noindent
where $R^\mu_\nu$ is the Ricci tensor, and the last term is the tail term, which consists of a non-local integral over the entire lifetime of the particle \citep{RRtailterm}. 

Since in neutron star magnetospheres, the dynamics are always dominated by the external electromagnetic fields, the tail term can be neglected \citep{RRtailtermneglect2,RRtailtermneglect1,jsantostail2,RRtailtermneglectstuchlik}. Furthermore, the metric that describes rotating neutron stars is the Hartle-Thorne metric, which is a vacuum solution of the Einstein field equations, regardless of whether the slow rotation approximation is used. Therefore, the Ricci tensor vanishes everywhere outside the star, and our description of radiation reaction reduces to the curved-spacetime generalization of the Landau-Lifshitz model:
\begin{align}
    \label{eq:curvedspaceLLRR}
    F_{RR}^\mu  = \frac{2q^3}{3m} &\bigg[ (\partial_\alpha F^{\mu}_{\nu} + \Gamma^\mu_{\delta \alpha} F^{\delta}_{\nu} -  \Gamma^\delta_{\nu \alpha} F^{\mu}_{\delta}) u^\nu u^\alpha \\ & \ \ + \frac{q}{m} ( F^{\mu\nu} F_{\nu\alpha} u^\alpha + F^{\alpha \beta} F_{\beta\nu} u_\alpha u^\nu u^\mu) \bigg] \ \nonumber .
\end{align}

The term that depends on the partial derivatives of the field tensor, $\partial_\alpha F^{\mu}_{\nu}$, contains no corrections due to spacetime curvature, and can thus be neglected as in flat spacetime \citep{RRderivativeneglect,marijaRR}.
Finally, the three-dimensional version of the general-relativistic reduced Landau-Lifshitz force is obtained by performing the 3+1 decomposition of equation (\ref{eq:curvedspaceLLRR}). The 3+1 split of the terms that do not involve curved-spacetime quantities simply yields the three-dimensional formulation of the Landau-Lifshitz model (\ref{eq:LLRR3D}), while the derivation of the 3+1 split of the gravitational contribution $\bvec{F}_{RR}^{(G)}$ is much more involved. For any generic stationary metric, expressed in Boyer-Lindquist coordinates, whose only non-zero diagonal entries are $g_{t\phi}$, this contribution is given by equation (\ref{eq:LLRRG3D}) \citep{jjthesis}.

The characteristic magnitude of the curved spacetime contribution can be estimated by comparison with the flat-spacetime formula. Since both the Christoffel symbols and the gravitational acceleration are first derivatives of the metric tensor, their magnitude is similar. Identifying the magnetic field as the dominant contribution, we find:

\begin{equation}
    \label{eq:RRratiomag}
    \frac{\left|\bvec{F}_{RR}^{(G)}\right|}{\left|\bvec{F}_{RR}^{(LL)}\right|} \sim \frac{\gamma |q|^3 v^2 g B/m}{\gamma^2|q|^4 v^3 B^2/m^2} = \frac{m g}{\gamma |q| v B} \ .
\end{equation}

\noindent

Given that the electromagnetic force is much stronger than the gravitational force, this ratio is generally small. However, since the gravitational acceleration scales with $1/r^2$ and the magnetic field with $1/r^3$, for a dipolar field, the contribution of the curved-spacetime corrections is proportional to the radial distance to the star.
For this reason, even though the gravitational term is negligible near the stellar surface, its contribution to the RR force can become significant at higher distances from the star, or when modelling longer evolutions of the system, as small phase-space deviations accumulate.

\section{Support equations for Section 3.1}
\label{app:math_section3}

For the sake of completeness and in order not to clutter the main text, we provide here some relations missing from Section \ref{subsec:spirals}.
The quantity $\bvec{p}_0(\bvec{p},t)$ is obtained by inverting equations \eqref{eq:approximate_motion_solution}, which forms a smooth map from $\{ \bvec{p},t \}$-space to $\{ \bvec{p}_0 \}$-space,
\begin{subequations}\label{eq:approximate_motion_solution_inverse}
\begin{align}
\begin{split}
p_{\perp 1 0} &= \frac{m c}{p' \vphantom{p}_{\!\!\perp}} \frac{p_{\perp 1} \cos \psi + p' \vphantom{p}_{\!\!\perp 2} \sin \psi}{\sinh \Big( \textrm{arsinh} \big( \frac{m c}{p' \vphantom{p}_{\!\!\perp}} \big) - \frac{\gamma' \vphantom{p}_{\!\!\perp}}{\gamma'} \omega_{RR} t \Big)} \ ,
\end{split}
\\[2pt]
\begin{split}
p' \vphantom{p}_{\!\!\perp 2 0} &= \frac{m c}{p' \vphantom{p}_{\!\!\perp}} \frac{p' \vphantom{p}_{\!\!\perp 2} \cos \psi  - p_{\perp 1} \sin \psi}{\sinh \Big( \textrm{arsinh} \big( \frac{m c}{p' \vphantom{p}_{\!\!\perp}} \big) - \frac{\gamma' \vphantom{p}_{\!\!\perp}}{\gamma'} \omega_{RR} t \Big)} \ ,
\end{split}
\\[2pt]
\begin{split}
p_{\parallel 0} &= \frac{1}{\gamma' \vphantom{p}_{\!\!\perp}} \frac{p_{\parallel}}{\tanh \Big( \textrm{artanh} \big( \frac{1}{\gamma' \vphantom{p}_{\!\!\perp}} \big) - \frac{\gamma' \vphantom{p}_{\!\!\perp}}{\gamma'} \omega_{RR} t \Big)} \ ,
\end{split}
\\[2pt]
\begin{split}
\gamma' \vphantom{p}_{\!0} &= \frac{1}{\gamma' \vphantom{p}_{\!\!\perp}} \frac{\gamma'}{\tanh \Big( \textrm{artanh} \big( \frac{1}{\gamma' \vphantom{p}_{\!\!\perp}} \big) - \frac{\gamma' \vphantom{p}_{\!\!\perp}}{\gamma'} \omega_{RR} t \Big)} \ .
\end{split}
\end{align}
\end{subequations}
All quantities are defined in the main text.
In addition, the gyrophase in equation \eqref{eq:gyrophase}, when written in terms of $\bvec{p}$, becomes:
\begin{equation}
\!\!\!\psi = \frac{\omega_c}{\omega_{RR}} \! \ln \!\!\, \left[ \gamma' \vphantom{p}_{\!\!\perp} \frac{m c}{p' \vphantom{p}_{\!\!\perp}} \! \left\{ \cosh \! \left( \textrm{arcosh} \! \left( \gamma' \vphantom{p}_{\!\!\perp} \frac{m c}{p' \vphantom{p}_{\!\!\perp}} \right) - \frac{\gamma' \vphantom{p}_{\!\!\perp}}{\gamma'} \omega_{RR} t \right) \right\}^{-1} \right] \! \ .
\end{equation}

\section{General-relativistic particle pusher}
\label{app:ppusher}

A particle pusher is a numerical algorithm designed to advance the positions and momenta of an ensemble of particles according to the forces that act upon them. As in curved spacetime coordinates are generally curved, we define all quantities, except the position vector, in a local orthonormal basis $\{ \bvec{e}_{\hat{i}} \}$ which satisfies $\bvec{e}_{\hat{i}} \cdot \bvec{e}_{\hat{j}} = \delta_{ij}$ and $\bvec{e}_{\hat{i}} \times \bvec{e}_{\hat{j}} = \varepsilon_{\hat{i}\hat{j}\hat{k}} \bvec{e}_{\hat{k}}$. The orthonormal basis vectors are thus defined by $\bvec{e}_{\hat{i}} = \frac{\bvec{\partial}_i}{h_i}$, where $\{\bvec{e}_i\} = \{\bvec{\partial}_i\}$ is the canonical basis, $h_i = \sqrt{g_{ii}}$ are the Lamé coefficients \citep{arfken} and no summation is implied in all equation terms containing $h_i$. 

If the metric tensor is spatially diagonal, which is the case in this work, the orthonormal basis expresses the physical components of tensors, since for any generic component index we have $\bvec{v} = v^i \bvec{e}_i = v^{\hat{i}} \bvec{e}_{\hat{i}}$. The orthonormal covariant and contravariant tensor components are then respectively given by:
\begin{equation}
    \label{eq:apporthbasisvec}
     v_{\hat{i}} = \frac{v_i}{h_i} \ \ , \ \ v^{\hat{i}} = h_i v^i \ .
\end{equation}

The use of the orthonormal basis greatly simplifies the manipulation of quantities in curvilinear coordinates, because the orthonormal frame is locally Cartesian. As such, the dot and cross products of spatial vectors reduce to their familiar Cartesian form, even when done in non-Cartesian coordinates:
\begin{align}
    \label{eq:appdotorth}
    & \bvec{A} \cdot \bvec{B} = \gamma_{ij} A^{i} B^{j} = \delta_{ij} A^{\hat{i}} B^{\hat{j}} = A^{\hat{1}} B^{\hat{1}} + A^{\hat{2}} B^{\hat{2}} + A^{\hat{3}} B^{\hat{3}} \ ,
\end{align}
\begin{align}
    \label{eq:appcrossorth}
    \bvec{A} \times \bvec{B} &= \delta^{ij} \varepsilon_{\hat{j}\hat{k}\hat{l}} A^{\hat{k}} B^{\hat{l}} \bvec{e}_{\hat{i}} \\ 
    &= (A^{\hat{2}} B^{\hat{3}} - A^{\hat{3}} B^{\hat{2}}, A^{\hat{3}} B^{\hat{1}} - A^{\hat{1}} B^{\hat{3}}, A^{\hat{1}} B^{\hat{2}} - A^{\hat{2}} B^{\hat{1}}) \ \nonumber .
\end{align}

\medskip 

Our particle pusher evolves the contravariant position and the orthonormalized contravariant spatial part of the four-velocity, $\bvec{w}=\gamma \bvec{v}$, which does not depend on the particle position. This scheme preserves numerical stability for particles with a very high Lorentz factor and provides a numerically convenient formulation in curved spacetime, by absorbing the off-diagonal vector product components. 

Writing the momentum equation of motion in terms of orthonormalized components introduces an additional fictitious force \cite{osirisgr}
\begin{equation}
    \label{eq:appcoordforce}
     \bvec{F}_{\text{coord}} \equiv -\frac{d \bvec{e}_{\hat{i}}}{dt} w^{\hat{i}} = -\overset{j \neq i}{\Gamma_{\hat{j}k}^{\hat{i}}} w^{\hat{j}} \frac{dx^k}{dt} \bvec{e}_{\hat{i}} \ ,
\end{equation}

\noindent
which is caused by the variation of the orthonormal basis. Finally, the equations of motion (\ref{eq:3+1eqmom}) in orthonormalized momentum components for a charged particle in curved spacetime, subject to radiation reaction, and written in the 3+1 formalism \cite{3+1EOM2,osirisgr} read:
\begin{align}
    \label{eq:appdxdt}
    & \frac{d x^i}{dt} = \frac{\alpha}{\gamma} \frac{w^{\hat{i}}}{h_i} - \frac{\beta^{\hat{i}}}{h_i} \ ,
\end{align}
\begin{align}
    \label{eq:appdwdt}
    \frac{d w^{\hat{i}}}{dt} &= \alpha \gamma g^{\hat{i}} + \alpha \delta^{ij} H_{\hat{j}\hat{k}} w^{\hat{k}} - \overset{j \neq i}{\Gamma_{\hat{j}k}^{\hat{i}}} w^{\hat{j}} \frac{dx^k}{dt} \\ &+ \frac{\alpha q}{m} \left(E^{\hat{i}} + \delta^{ij} \varepsilon_{\hat{j}\hat{k}\hat{l}} \frac{w^{\hat{k}}}{\gamma} B^{\hat{l}} \right) + \frac{\alpha}{m} F_{RR}^{\hat{i}} \ \nonumber .
\end{align}

\noindent

The time integration of equations (\ref{eq:appdxdt}) and (\ref{eq:appdwdt}) is performed using an explicit 6\textsuperscript{th} order Runge–Kutta method \citep{RK6} with a fixed timestep interval of 50 points in each gyroperiod, by default, so as to adequately resolve the particle gyro-orbits. As this formulation is fully relativistic, the algorithm remains valid for hyper-relativistic trajectories and accurately captures high-Lorentz-factor particle dynamics.


\bsp	
\label{lastpage}
\end{document}